# THE EQUATIONS OF THE IDEAL LATCHES

SERBAN E. VLAD

Oradea City Hall, Piata Unirii Nr. 1, 410100, Oradea, Romania

[www.geocities.com/serban_e_vlad](www.geocities.com/serban_e_vlad), serban_e_vlad@yahoo.com

ABSTRACT

*We present the equations that model several types of latches and flip-flops. The circuits are supposed to be ideal (i.e. non-inertial).*

**Contents** : 1. Introduction, 2. Signals, 3. Latches, the general equation, 4. C element, 5. RS latch, 6. Clocked RS latch, 7. D latch, 8. Edge triggered RS flip-flop, 9. D flip-flop, 10. JK flip-flop, 11. T flip-flop, Appendix

## 1. Introduction

The latches are simple circuits with feedback from the digital electrical engineering. We have included in our work the C element of Muller, the RS latch, the clocked RS latch, the D latch and also circuits containing two interconnected latches: the edge triggered RS flip-flop, the D flip-flop, the JK flip-flop, the T flip-flop.

The purpose of this study is to model with equations the previous circuits, considered to be ideal, i.e. non-inertial. In the case of the RS latch for example, this refers to the system

$$\overline{Q(t-0)} \cdot Q(t) = \overline{Q(t-0)} \cdot S(t)$$
$$Q(t-0) \cdot \overline{Q(t)} = Q(t-0) \cdot R(t)$$
$$R(t) \cdot S(t) = 0$$

where $R, S, Q : \boldsymbol{R} \to \{0, 1\}$ are signals. One of the inertial versions of this system is

$$\overline{Q(t-0)} \cdot Q(t) = \overline{Q(t-0)} \cdot \bigcap_{\xi \in [t-d_r, t]} S(\xi)$$
$$Q(t-0) \cdot \overline{Q(t)} = Q(t-0) \cdot \bigcap_{\xi \in [t-d_f, t]} R(\xi)$$
$$\bigcap_{\xi \in [t-d_f, t]} R(\xi) \cdot \bigcap_{\xi \in [t-d_r, t]} S(\xi) = 0$$

with $d_r \geq 0, d_f \geq 0$ inertial parameters. The above equations highlight the fact that 'short' 1 values of $R, S$ cannot influence the behavior of the circuit.

The technique of analysis is the pseudoboolean differential calculus.

The internet address that we have indicated at the bibliography created some order in the present thoughts, but the latches are very well known to the electrical engineers and the bibliography concerning them is rich.

## 2. Signals

We note with $B = \{0,1\}$ the Boole algebra with two elements and $\chi_A : R \to B$ is the notation of the characteristic function of the set $A \subset R$.

The (electrical) signals are the functions $x : R \to B$ having the property that some unbounded sequence $0 \le t_0 < t_1 < t_2 < ...$ exists s.t. $\forall t \in R$,

$$x(t) = x(-1) \cdot \chi_{(-\infty, t_0)}(t) \oplus x(t_0) \cdot \chi_{[t_0, t_1)}(t) \oplus x(t_1) \cdot \chi_{[t_1, t_2)}(t) \oplus ...$$

The signals have the property of existence of the left limit:
$$\forall t, \exists \varepsilon > 0, \forall \xi \in (t - \varepsilon, t), x(\xi) = x(t - 0)$$

even if $x(t-0)$ is not a signal

$$x(t-0) = x(-1) \cdot \chi_{(-\infty, t_0]}(t) \oplus x(t_0) \cdot \chi_{(t_0, t_1]}(t) \oplus x(t_1) \cdot \chi_{(t_1, t_2]}(t) \oplus ...$$

For each signal $x$, the left semi-derivatives $\overline{x(t-0)} \cdot x(t), x(t-0) \cdot \overline{x(t)}$ are defined:

$$\overline{x(t-0)} \cdot x(t) = \overline{x(-1)} \cdot x(t_0) \cdot \chi_{\{t_0\}}(t) \oplus \overline{x(t_0)} \cdot x(t_1) \cdot \chi_{\{t_1\}}(t) \oplus \overline{x(t_1)} \cdot x(t_2) \cdot \chi_{\{t_2\}}(t) \oplus ...$$

$$x(t-0) \cdot \overline{x(t)} = x(-1) \cdot \overline{x(t_0)} \cdot \chi_{\{t_0\}}(t) \oplus x(t_0) \cdot \overline{x(t_1)} \cdot \chi_{\{t_1\}}(t) \oplus x(t_1) \cdot \overline{x(t_2)} \cdot \chi_{\{t_2\}}(t) \oplus ...$$

showing the discrete time instances $t_0, t_1, t_2, ...$ when $x$ may switch from 0 to 1, respectively from 1 to 0.

Let $x, y$ two signals. We shall use the fact that the left limit of $x(t-0)$, $\overline{x(t)}$, $x(t) \cdot y(t)$, $x(t) \cup y(t)$ is respectively $x(t-0)$, $\overline{x(t-0)}$, $x(t-0) \cdot y(t-0)$, $x(t-0) \cup y(t-0)$.

## 3. Latches, the general equation

The *equations of the latches* consist in the next system

$$\overline{x(t-0)} \cdot x(t) = \overline{x(t-0)} \cdot u(t)$$
$$x(t-0) \cdot \overline{x(t)} = x(t-0) \cdot v(t) \qquad (1)$$
$$u(t) \cdot v(t) = 0$$

where $u, v, x$ are signals and $x$ is the unknown. The last equation of the system is called the *admissibility condition* (of the inputs).

In order to solve system (1) we associate to the functions $u, v$ the next sets $U_{2k}, V_{2k+1}$ and respectively numbers $t_k$ :

$$U_0 = \{t \mid \overline{u(t-0)} \cdot u(t) = 1\}, \qquad t_0 = \min U_0$$
$$V_1 = \{t \mid \overline{v(t-0)} \cdot v(t) = 1, t > t_0\}, \qquad t_1 = \min V_1$$
$$U_2 = \{t \mid \overline{u(t-0)} \cdot u(t) = 1, t > t_1\}, \qquad t_2 = \min U_2$$
$$V_3 = \{t \mid \overline{v(t-0)} \cdot v(t) = 1, t > t_2\}, \qquad t_3 = \min V_3$$
$$...$$

and the next inclusions, respectively inequalities are true

$$U_0 \supset U_2 \supset U_4 \supset ... \qquad V_1 \supset V_3 \supset V_5 \supset ...$$
$$0 \le t_0 < t_1 < t_2 < ...$$

For each of $U_{2k} (V_{2k+1})$ we have the possibilities:

- it is empty. Then $t_{2k} (t_{2k+1})$ is undefined and all $U_{2k}, V_{2k+1}, t_k$ of higher rank are undefined

- it is non-empty, finite or infinite. $t_{2k}$ ($t_{2k+1}$) is defined.

If $U_{2k}$ ($V_{2k+1}$) are defined for all $k \in N$, then the sequence $(t_k)$ is unbounded.

A similar discussion is related with the sets $V'_{2k}, U'_{2k+1}$ and respectively the numbers $t'_k$:

$$V'_0 = \{t \mid \overline{v(t-0)} \cdot v(t) = 1\}, \qquad t'_0 = \min V'_0$$
$$U'_1 = \{t \mid \overline{u(t-0)} \cdot u(t) = 1, t > t'_0\}, \qquad t'_1 = \min U'_1$$
$$V'_2 = \{t \mid \overline{v(t-0)} \cdot v(t) = 1, t > t'_1\}, \qquad t'_2 = \min V'_2$$
$$U'_3 = \{t \mid \overline{u(t-0)} \cdot u(t) = 1, t > t'_2\}, \qquad t'_3 = \min U'_3$$

...

For solving the system (1), we observe that the unbounded sequence $0 \le t''_0 < t''_1 < t''_2 < ...$ exists with the property that $u, v, x$ are constant in each of the intervals $(-\infty, t''_0), [t''_0, t''_1), [t''_1, t''_2), ...$ where the first two equations of (1) take one of the forms

$$\overline{x(t-0)} \cdot x(t) = \overline{x(t-0)} \qquad (2)$$
$$x(t-0) \cdot \overline{x(t)} = 0$$

$$\overline{x(t-0)} \cdot x(t) = 0 \qquad (3)$$
$$x(t-0) \cdot \overline{x(t)} = x(t-0)$$

$$\overline{x(t-0)} \cdot x(t) = 0 \qquad (4)$$
$$x(t-0) \cdot \overline{x(t)} = 0$$

as $u(t), v(t)$ are equal with 1,0; 0,1; 0,0 in those intervals. The solutions were written in the next table:

| | $eq.(2)$ $u(t)=1, v(t)=0$ | $eq.(3)$ $u(t)=0, v(t)=1$ | $eq.(4)$ $u(t)=0, v(t)=0$ |
|---|---|---|---|
| $t \in (-\infty, t''_0)$ | $x(t)=1$ | $x(t)=0$ | $x(t)=0$ $x(t)=1$ |
| $t \in [t''_k, t''_{k+1})$ | $x(t)=1$ | $x(t)=0$ | $x(t)=x(t''_k-0)$ |

Table 1

We observe furthermore (see the Appendix) that (1) is equivalent with the equation

$$x(t) \cdot u(t) \cdot \overline{v(t)} \cup \overline{x(t)} \cdot \overline{u(t)} \cdot v(t) \cup (\overline{x(t-0)} \cdot \overline{x(t)} \cup x(t-0) \cdot x(t)) \cdot \overline{u(t)} \cdot \overline{v(t)} = 1 \quad (5)$$

that contains three exclusive possibilities: $x(t) \cdot u(t) \cdot \overline{v(t)} = 1$, $\overline{x(t)} \cdot \overline{u(t)} \cdot v(t) = 1$ respectively $(\overline{x(t-0)} \cdot \overline{x(t)} \cup x(t-0) \cdot x(t)) \cdot \overline{u(t)} \cdot \overline{v(t)} = 1$ equivalent with (2), (3), (4).

We solve the system (1).

<u>Case a)</u> $u(0-0)=0, v(0-0)=0$

$x(0-0)=0$ and $x(0-0)=1$ are both possible. In order to make a distinction between the two solutions of (1) corresponding to the initial value 0, respectively the initial value 1 we shall note them with $x$, respectively with $x'$.

a.i) $x(0-0) = 0$

a.i.1) $U_0 = \varnothing$

the solution of (1) is $x(t) = 0$

a.i.2) $U_0 \neq \varnothing$

and $\exists \varepsilon > 0, x(t) = \chi_{[t_0,\infty)}(t)$ for $t < t_0 + \mathbf{e}$. This fact results by solving (4) for $t < t_0$ and then (2) followed perhaps by a finite sequence of (4), (2), (4),... in some interval $[t_0, t_0 + \mathbf{e})$. Furthermore:

a.i.2.1) $V_1 = \varnothing$

the solution of (1) is $x(t) = \chi_{[t_0,\infty)}(t)$

a.i.2.2) $V_1 \neq \varnothing$

and $\exists \varepsilon > 0, x(t) = \chi_{[t_0,t_1)}(t)$ for $t < t_1 + \varepsilon$. In some interval $[t_1, t_1 + \varepsilon)$ we solved (3) followed perhaps by a finite sequence of (4), (3), (4),...

a.i.2.2.1) $U_2 = \varnothing$

the solution of (1) is $x(t) = \chi_{[t_0,t_1)}(t)$

a.i.2.2.2) $U_2 \neq \varnothing$

and $\exists \varepsilon > 0, x(t) = \chi_{[t_0,t_1)}(t) \oplus \chi_{[t_2,\infty)}(t)$ for $t < t_2 + \varepsilon$.

a.i.2.2.2.1) $V_3 = \varnothing$

the solution of (1) is $x(t) = \chi_{[t_0,t_1)}(t) \oplus \chi_{[t_2,\infty)}(t)$

a.i.2.2.2.2) $V_3 \neq \varnothing$

$\ldots$

a.ii) $x'(0-0) = 1$

a.ii.1) $V_0' = \varnothing$

the solution of (1) is $x'(t) = 1$

a.ii.2) $V_0' \neq \varnothing$

$\exists \varepsilon > 0, x'(t) = \chi_{(-\infty,t_0')}(t)$ for all $t < t_0' + \varepsilon$

a.ii.2.1) $U_1' = \varnothing$

the solution of (1) is $x'(t) = \chi_{(-\infty,t_0')}(t)$

a.ii.2.2) $U_1' \neq \varnothing$

$\exists \varepsilon > 0, x'(t) = \chi_{(-\infty,t_0')}(t) \oplus \chi_{[t_1',\infty)}(t)$ for all $t < t_1' + \varepsilon$

$\ldots$

We have drawn in Fig 1 and Fig 2 the solutions $x, x'$ corresponding to Case a) in the situation when $t_0 < t_0'$, respectively when $t_0 > t_0'$ (the equality $t_0 = t_0'$ is impossible, because it implies $u(t_0) = v(t_0') = 1$, contradiction with (1)). We observe the fact that $x_{|[t_0,\infty)} = x'_{|[t_0,\infty)}$, respectively $x_{|[t_0',\infty)}' = x'_{|[t_0',\infty)}$ thus after the first common value of the (distinct) solutions $x, x'$ they coincide.

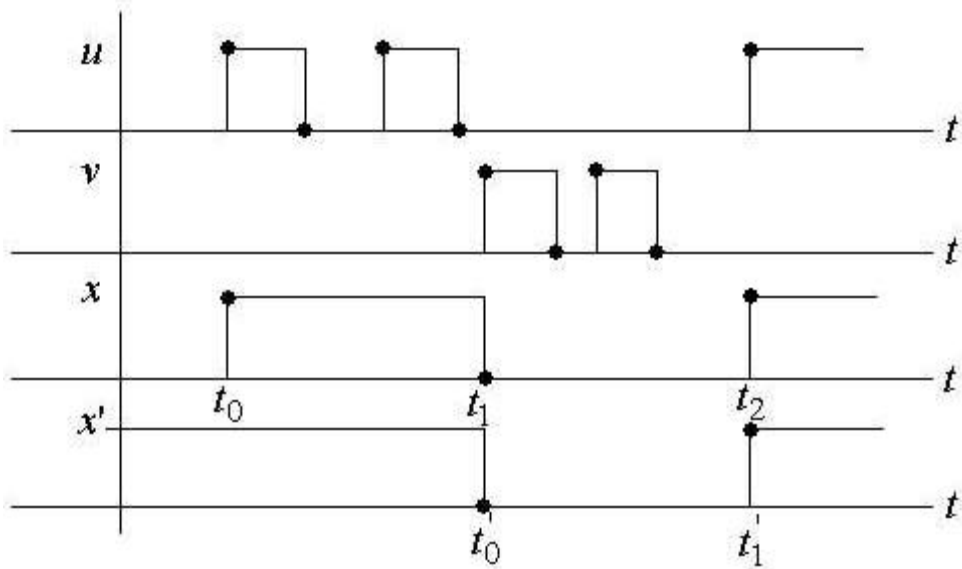

Fig 1

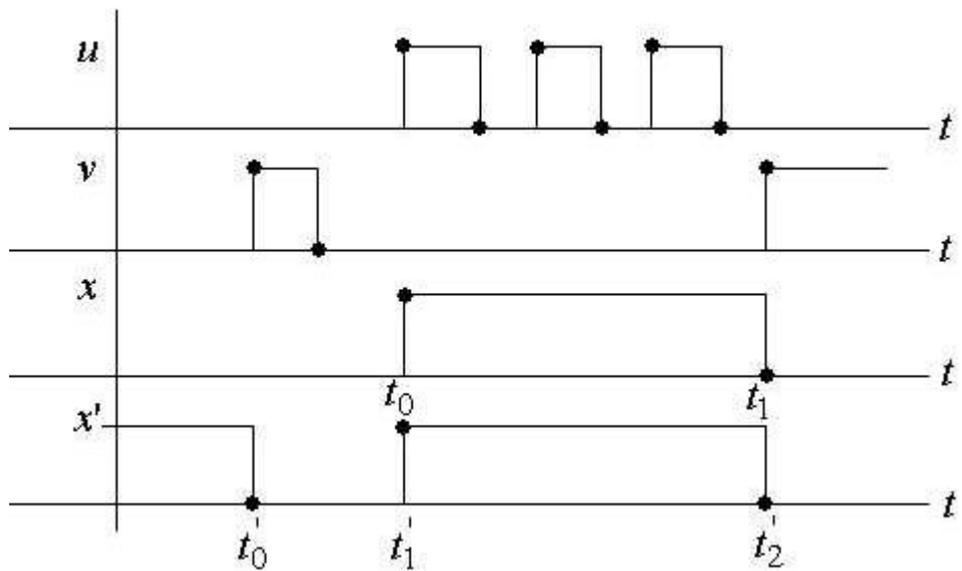

Fig 2

<u>Case b)</u> $u(0-0) = 1, v(0-0) = 0$

the only possibility is $x(0-0) = 1$

        b.1) $V_0^{'} = \varnothing$

the solution of (1) is $x(t) = 1$

        b.2) $V_0^{'} \neq \varnothing$

$\exists \varepsilon > 0, x(t) = \chi_{(-\infty, t_0^{'})}(t)$ for all $t < t_0^{'} + \varepsilon$

                    ...

<u>Case c)</u> $u(0-0) = 0, v(0-0) = 1$

the only possibility is $x(0-0) = 0$

        c.1) $U_0 = \varnothing$

the solution of (1) is $x(t) = 0$

## 4. C element

We call the *equations of the C element of Muller* any of the next equivalent statements:

$$\overline{x(t-0)} \cdot x(t) = \overline{x(t-0)} \cdot u(t) \cdot v(t) \qquad (1)$$

$$x(t-0) \cdot \overline{x(t)} = x(t-0) \cdot \overline{u(t)} \cdot \overline{v(t)}$$

and respectively

$$x(t) \cdot u(t) \cdot v(t) \cup \overline{x(t)} \cdot \overline{u(t)} \cdot \overline{v(t)} \cup \qquad (2)$$

$$\cup \left( \overline{x(t-0)} \cdot \overline{x(t)} \cup x(t-0) \cdot x(t) \right) \cdot \left( \overline{u(t)} \cdot v(t) \cup u(t) \cdot \overline{v(t)} \right) = 1$$

where $u, v, x$ are signals, the first two called *inputs* and the last – *state*.

Equations (1), (2) are the equations of a latch 3 (1), 3 (5) where $u(t)$ is replaced by $u(t) \cdot v(t)$ and $v(t)$ is replaced by $\overline{u(t)} \cdot \overline{v(t)}$. It is observed the satisfaction of the admissibility condition of the inputs.

The circuit called the C element of Muller and its symbol were drawn bellow:

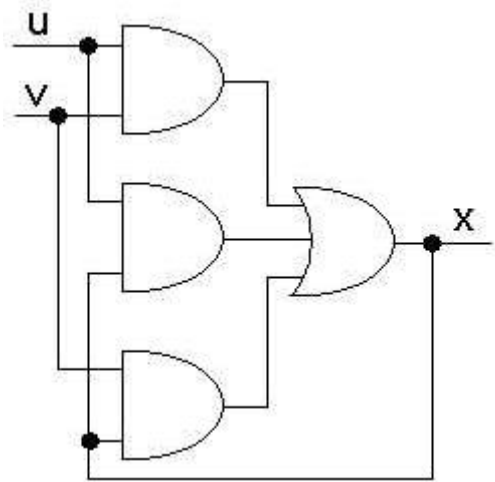

a)

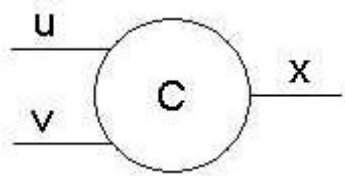

b)

Fig 3

The analysis of (2) is obvious: $x(t)$ is 1 if $u(t) = v(t) = 1$, $x(t)$ is 0 if $u(t) = v(t) = 0$ and $x(t) = x(t-0), x(t)$ keeps its previous value otherwise.

The general form of equation (1) for $n$ inputs $u_1,...,u_n$ is:

$$\overline{x(t-0)} \cdot x(t) = \overline{x(t-0)} \cdot u_1(t) \cdot ... \cdot u_n(t)$$

$$x(t-0) \cdot \overline{x(t)} = x(t-0) \cdot \overline{u_1(t)} \cdot \ldots \cdot \overline{u_n(t)}$$

and equation (2) is generalized like this:

$$x(t) \cdot u_1(t) \cdot \ldots \cdot u_n(t) \cup \overline{x(t)} \cdot \overline{u_1(t)} \cdot \ldots \cdot \overline{u_n(t)} \cup$$
$$\cup (\overline{x(t-0)} \cdot \overline{x(t)} \cup x(t-0) \cdot x(t)) \cdot \overline{u_1(t)} \cdot \ldots \cdot \overline{u_n(t)} \cdot (u_1(t) \cup \ldots \cup u_n(t)) = 1$$

## 5. RS latch

We call the *equations of the RS latch* any of the equivalent statements

$$\overline{Q(t-0)} \cdot Q(t) = \overline{Q(t-0)} \cdot S(t)$$
$$Q(t-0) \cdot \overline{Q(t)} = Q(t-0) \cdot R(t) \qquad (1)$$
$$R(t) \cdot S(t) = 0$$

and respectively

$$Q(t) \cdot \overline{R(t)} \cdot S(t) \cup \overline{Q(t)} \cdot R(t) \cdot \overline{S(t)} \cup (\overline{Q(t-0)} \cdot \overline{Q(t)} \cup Q(t-0) \cdot Q(t)) \cdot \overline{R(t)} \cdot \overline{S(t)} = 1 \quad (2)$$

In (1), (2) $R, S, Q$ are signals. $R, S$ are called inputs: the *reset input* and the *set input* and $Q$ is the *state*, the unknown relative to which the equations are solved.

These equations coincide with 3 (1) and 3 (5) but the notations are different and traditional.

We have drawn below the circuit called RS latch and its symbol.

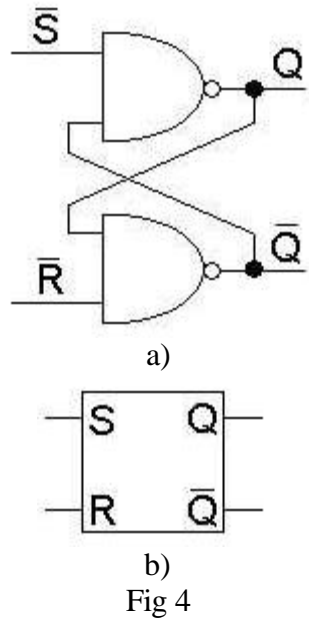

a)

b)

Fig 4

We conclude the things that were discussed in section 3 by the next statements related with equation (2). At the RS latch, $Q(t) = 1$ if $R(t) = 0, S(t) = 1$; $Q(t) = 0$ if $R(t) = 1, S(t) = 0$; and $Q(t) = Q(t-0)$, $Q$ keeps its previous value if $R(t) = 0, S(t) = 0$.

## 6. Clocked RS latch

The equivalent statements

$$\overline{Q(t-0)} \cdot Q(t) = \overline{Q(t-0)} \cdot S(t) \cdot C(t)$$
$$Q(t-0) \cdot \overline{Q(t)} = Q(t-0) \cdot R(t) \cdot C(t) \qquad (1)$$
$$R(t) \cdot S(t) \cdot C(t) = 0$$

and
$$C(t) \cdot (Q(t) \cdot \overline{R(t)} \cdot S(t) \cup \overline{Q(t)} \cdot R(t) \cdot \overline{S(t)} \cup$$
$$(\overline{Q(t-0) \cdot \overline{Q(t)}} \cup Q(t-0) \cdot Q(t)) \cdot \overline{R(t)} \cdot \overline{S(t)}) \cup \qquad (2)$$
$$\overline{C(t)} \cdot (\overline{Q(t-0) \cdot \overline{Q(t)}} \cup Q(t-0) \cdot Q(t)) = 1$$

are called the *equations of the clocked RS latch*. $R, S, C, Q$ are signals: the *reset*, the *set* and the *clock* input, respectively the *state*.

The equations (1), (2) result from 3 (1) and 3 (5) where $u(t) = S(t) \cdot C(t)$, $v(t) = R(t) \cdot C(t)$. The circuit called the clocked RS latch and the symbol of this circuit were drawn below

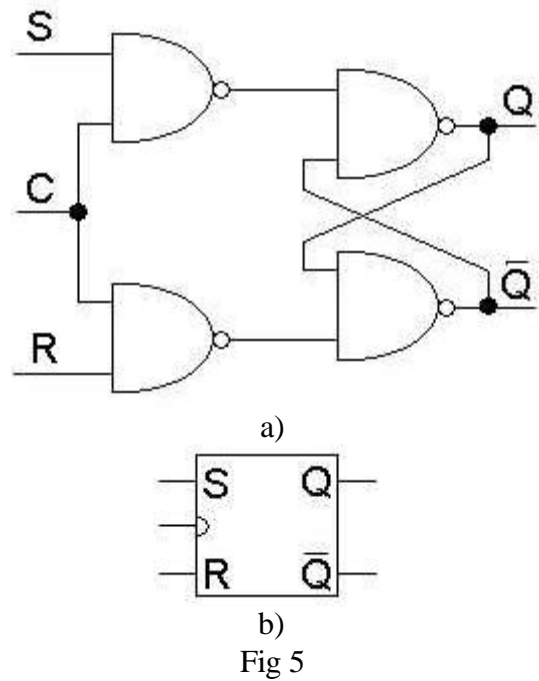

a)

b)

Fig 5

The work of the circuit results from equation (2):

<u>Case</u> $C(t) = 1$ We get
$$Q(t) \cdot \overline{R(t)} \cdot S(t) \cup \overline{Q(t)} \cdot R(t) \cdot \overline{S(t)} \cup (\overline{Q(t-0) \cdot \overline{Q(t)}} \cup Q(t-0) \cdot Q(t)) \cdot \overline{R(t)} \cdot \overline{S(t)} = 1$$
i.e. the equation of the RS latch 5 (2)

<u>Case</u> $\overline{C(t)} = 1$, expressing the validity of
$$\overline{Q(t-0) \cdot \overline{Q(t)}} \cup Q(t-0) \cdot Q(t) = 1$$

shows the constancy of $Q$: $Q(t) = Q(t-0)$ is true in this interval.

The clocked RS latch behaves like an RS latch when $C(t) = 1$ and keeps the state constant $Q(t) = Q(t-0)$ when $C(t) = 0$, but these were obvious when looking at (1).

## 7. D latch

We call the *equations of the D latch* any of the next equivalent statements
$$\overline{Q(t-0)} \cdot Q(t) = \overline{Q(t-0)} \cdot D(t) \cdot C(t) \qquad (1)$$
$$Q(t-0) \cdot \overline{Q(t)} = Q(t-0) \cdot \overline{D(t)} \cdot C(t)$$

and respectively
$$C(t) \cdot (\overline{Q(t)} \cdot \overline{D(t)} \cup Q(t) \cdot D(t)) \cup \overline{C(t)} \cdot (\overline{Q(t-0) \cdot \overline{Q(t)}} \cup Q(t-0) \cdot Q(t)) = 1 \ (2)$$

$D, C, Q$ are signals: the *data input* $D$, the *clock input* $C$ and the *state* $Q$.

On one hand, from (1) it is seen the satisfaction of the admissibility condition of the inputs. And on the other hand (1), (2) result from the equations of the clocked RS latch 6 (1), 6 (2) where $R = \overline{S \cdot C}$ because

$$R(t) \cdot C(t) = \overline{S(t) \cdot C(t)} \cdot C(t) = (\overline{S(t)} \cup \overline{C(t)}) \cdot C(t) = \overline{S(t)} \cdot C(t)$$

and we have used the traditional notation $D$ for the data input, instead of $S$.

The D latch circuit and its symbol were drawn below

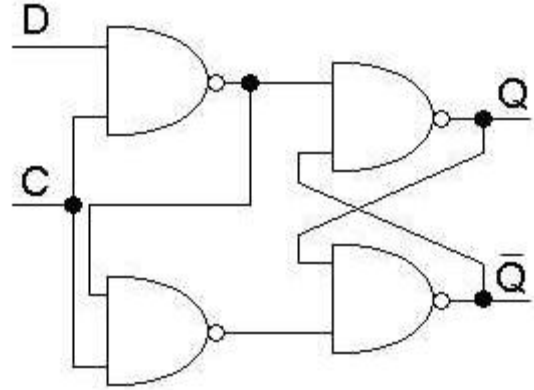

a)

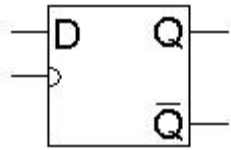

b)

Fig 6

We interpret the equations of the D latch now.

<u>Case</u> $C(t) = 1$

$$\overline{Q(t)} \cdot \overline{D(t)} \cup Q(t) \cdot D(t) = 1$$

$Q$ and $D$ coincide

<u>Case</u> $\overline{C(t)} = 1$

$$\overline{Q(t-0)} \cdot \overline{Q(t)} \cup Q(t-0) \cdot Q(t) = 1$$

$Q$ is constant.

When $C(t) = 1$, the D latch makes $Q(t) = D(t)$ and when $C(t) = 0$, $Q$ is constant.

## 8. Edge triggered RS flip-flop

Any of the equivalent statements

$$\overline{P(t-0)} \cdot P(t) = \overline{P(t-0)} \cdot S(t) \cdot C(t)$$
$$P(t-0) \cdot \overline{P(t)} = P(t-0) \cdot R(t) \cdot C(t)$$
$$R(t) \cdot S(t) \cdot C(t) = 0 \qquad\qquad (1)$$
$$\overline{Q(t-0)} \cdot Q(t) = \overline{Q(t-0)} \cdot \overline{P(t)} \cdot \overline{C(t)}$$
$$Q(t-0) \cdot \overline{Q(t)} = Q(t-0) \cdot \overline{P(t)} \cdot \overline{C(t)}$$

and respectively

$$C(t) \cdot (\overline{Q(t-0)} \cdot \overline{Q(t)} \cup Q(t-0) \cdot Q(t)) \cdot \qquad (2)$$

$$\cdot (P(t) \cdot \overline{R(t)} \cdot S(t) \cup \overline{P(t)} \cdot R(t) \cdot \overline{S(t)} \cup (\overline{P(t-0)} \cdot \overline{P(t)} \cup P(t-0) \cdot P(t)) \cdot \overline{R(t)} \cdot \overline{S(t)}) \cup$$

$$\overline{C(t)} \cdot (\overline{Q(t)} \cdot \overline{P(t-0)} \cdot \overline{P(t)} \cup Q(t) \cdot P(t-0) \cdot P(t)) = 1$$

is called the *equation of the edge triggered RS flip-flop*. $R, S, C, P, Q$ are signals: the *reset input* $R$, the *set input* $S$, the *clock input* $C$, the *next state* $P$ and the *state* $Q$.

In (1), (2) the signals $R, S, C, P$ and $P, \overline{C}, Q$ satisfy the equations of a clocked RS latch and of a D latch and (2) represents the term by term product of 6 (2) with 7 (2) written with these variables.

The two latches are called *master* and *slave*.

The edge triggered RS flip-flop circuit and its symbol are the following:

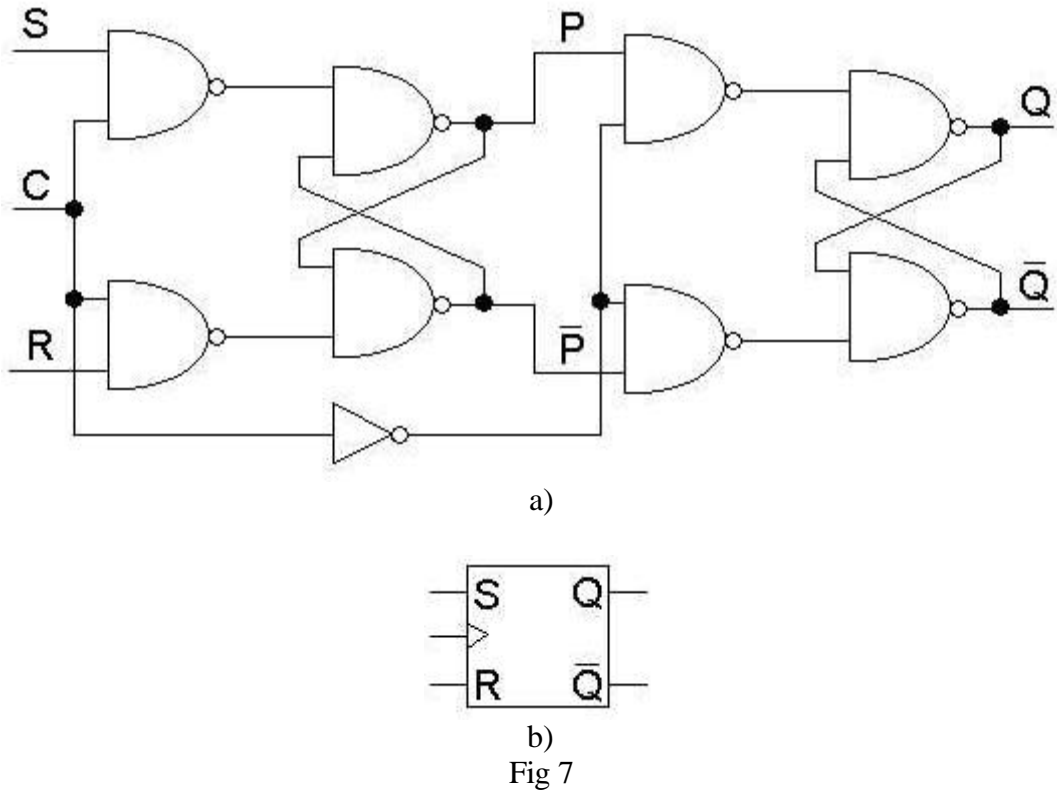

a)

b)

Fig 7

The equation (2) written at the left of $t$ is:

$$C(t-0) \cdot (P(t-0) \cdot \overline{R(t-0)} \cdot S(t-0) \cup \qquad (3)$$

$$\cup \overline{P(t-0)} \cdot R(t-0) \cdot \overline{S(t-0)} \cup \overline{R(t-0)} \cdot \overline{S(t-0)}) \cup$$

$$\overline{C(t-0)} \cdot (\overline{Q(t-0)} \cdot \overline{P(t-0)} \cup Q(t-0) \cdot P(t-0)) = 1$$

In the analysis of the circuit we have the next possibilities:

<u>Case</u> $\overline{C(t-0)} \cdot C(t) = 1$

$$\overline{Q(t-0)} \cdot \overline{P(t-0)} \cup Q(t-0) \cdot P(t-0) = 1$$

$$\overline{Q(t-0)} \cdot \overline{Q(t)} \cup Q(t-0) \cdot Q(t) = 1 \qquad (4)$$

$$P(t) \cdot \overline{R(t)} \cdot S(t) \cup \overline{P(t)} \cdot R(t) \cdot \overline{S(t)} \cup (\overline{P(t-0)} \cdot \overline{P(t)} \cup P(t-0) \cdot P(t)) \cdot \overline{R(t)} \cdot \overline{S(t)} = 1 \ (5)$$

i.e. $Q(t-0) = P(t-0) = Q(t)$ and on the other hand $R, S, P$ is an RS latch.

<u>Case</u> $C(t) = 1$

Only (4), (5) are true; $Q$ is constant and $R, S, P$ is an RS latch.

<u>Case</u> $C(t-0) \cdot \overline{C(t)} = 1$

$$P(t-0) \cdot \overline{R(t-0)} \cdot S(t-0) \cup \overline{P(t-0)} \cdot R(t-0) \cdot \overline{S(t-0)} \cup \overline{R(t-0)} \cdot \overline{S(t-0)} = 1 \qquad (6)$$

$$\overline{Q(t)} \cdot \overline{P(t-0)} \cdot \overline{P(t)} \cup Q(t) \cdot P(t-0) \cdot P(t) = 1 \qquad (7)$$

i.e. $R, S, P$ is an RS latch at the left of $t$, the value of $P(t-0)$ being transmitted to $P(t)$ and $Q(t)$.

<u>Case</u> $\overline{C(t)} = 1$

Only (7) is true, $P, Q$ are equal and constant.

   The name of edge triggered RS flip-flop refers to the fact that $Q(t)$ is constant at all time instances except $C(t-0) \cdot \overline{C(t)} = 1$, when $Q(t) = P(t-0) =$
$$= \begin{cases} 1, if \ R(t-0) = 0, S(t-0) = 1 \\ 0, if \ R(t-0) = 1, S(t-0) = 0 \end{cases}$$, this is the so called 'falling edge' of the clock input.

## 9. D flip-flop

We call the *equations of the D flip-flop* any of the next equivalent conditions:

$$\begin{aligned}
\overline{P(t-0)} \cdot P(t) &= \overline{P(t-0)} \cdot D(t) \cdot C(t) \\
P(t-0) \cdot \overline{P(t)} &= P(t-0) \cdot \overline{D(t)} \cdot C(t) \\
\overline{Q(t-0)} \cdot Q(t) &= \overline{Q(t-0)} \cdot P(t) \cdot \overline{C(t)} \\
Q(t-0) \cdot \overline{Q(t)} &= Q(t-0) \cdot \overline{P(t)} \cdot \overline{C(t)}
\end{aligned} \qquad (1)$$

and respectively

$$C(t) \cdot (\overline{Q(t-0)} \cdot \overline{Q(t)} \cup Q(t-0) \cdot Q(t)) \cdot (\overline{P(t)} \cdot \overline{D(t)} \cup P(t) \cdot D(t)) \cup \qquad (2)$$
$$\overline{C(t)} \cdot (\overline{Q(t)} \cdot \overline{P(t-0)} \cdot \overline{P(t)} \cup Q(t) \cdot P(t-0) \cdot P(t)) = 1$$

$D, C, P, Q$ are signals, called: the *data input D*, the *clock input C*, the *next state P* and the *state Q*.

   We observe that the equations of the $D$ flip-flop represent the special case of edge triggered $RS$ flip-flop when $R = \overline{S \cdot C}$ and $S$ was noted with $D$.

   The $D$ flip-flop circuit and its symbol are the next ones:

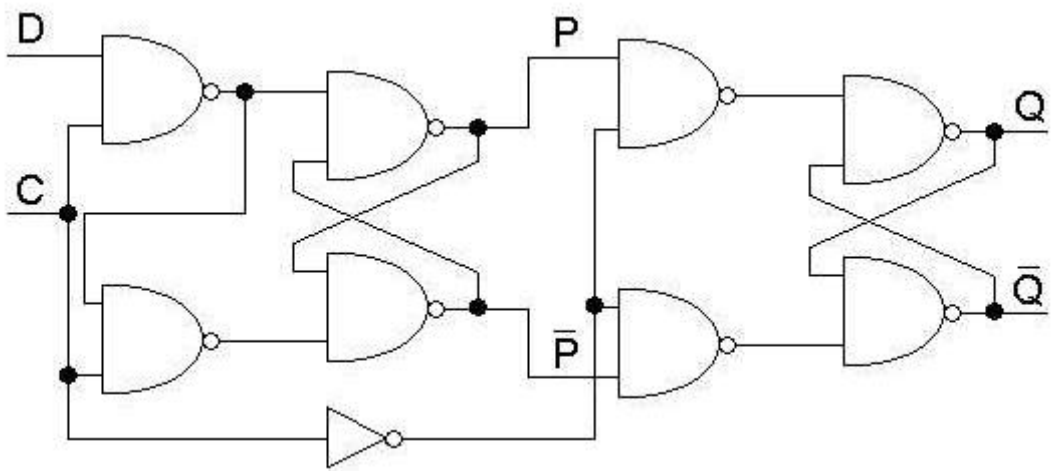

a)

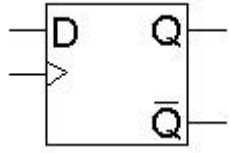

b)

Fig 8

We write equation (2) at the left of $t$:

$$C(t-0) \cdot (\overline{P(t-0)} \cdot \overline{D(t-0)} \cup P(t-0) \cdot D(t-0)) \cup$$
$$\overline{C(t-0)} \cdot (\overline{Q(t-0)} \cdot \overline{P(t-0)} \cup Q(t-0) \cdot P(t-0)) = 1$$

and we interpret the equations of the D flip-flop.

<u>Case</u> $\overline{C(t-0)} \cdot C(t) = 1$

$$(\overline{Q(t-0)} \cdot \overline{Q(t)} \cdot \overline{P(t-0)} \cup Q(t-0) \cdot Q(t) \cdot P(t-0)) \cdot (\overline{P(t)} \cdot \overline{D(t)} \cup P(t) \cdot D(t)) = 1$$

The state and the next state were equal, $Q$ keeps its value and the next state becomes equal with the input.

<u>Case</u> $C(t) = 1$

$$(\overline{Q(t-0)} \cdot \overline{Q(t)} \cup Q(t-0) \cdot Q(t)) \cdot (\overline{P(t)} \cdot \overline{D(t)} \cup P(t) \cdot D(t)) = 1$$

$Q$ keeps its value and $P$ is equal with $D$

<u>Case</u> $C(t-0) \cdot \overline{C(t)} = 1$

$$\overline{Q(t)} \cdot \overline{P(t-0)} \cdot \overline{P(t)} \cdot \overline{D(t-0)} \cup Q(t) \cdot P(t-0) \cdot P(t) \cdot D(t-0) = 1$$

the next state is equal with the previous value of the input and with the state; $t$ is a point of continuity of $P$

<u>Case</u> $\overline{C(t)} = 1$

$$\overline{Q(t)} \cdot \overline{P(t-0)} \cdot \overline{P(t)} \cup Q(t) \cdot P(t-0) \cdot P(t) = 1$$

the next state and the state have equal and constant values.

The D flip-flop has the state $Q$ constant except for the time instants when $C(t-0) \cdot \overline{C(t)} = 1$; then $Q(t) = D(t-0)$.

## 10. JK flip-flop

The equivalent statements

$$\overline{P(t-0)} \cdot P(t) = \overline{P(t-0)} \cdot J(t) \cdot \overline{Q(t)} \cdot C(t)$$
$$P(t-0) \cdot \overline{P(t)} = P(t-0) \cdot K(t) \cdot Q(t) \cdot C(t) \qquad (1)$$
$$\overline{Q(t-0)} \cdot Q(t) = \overline{Q(t-0)} \cdot P(t) \cdot \overline{C(t)}$$
$$Q(t-0) \cdot \overline{Q(t)} = Q(t-0) \cdot \overline{P(t)} \cdot \overline{C(t)}$$

and

$$C(t) \cdot (\overline{Q(t-0)} \cdot \overline{Q(t)} \cup Q(t-0) \cdot Q(t)) \cdot (P(t) \cdot J(t) \cdot \overline{Q(t)} \cup \overline{P(t)} \cdot K(t) \cdot Q(t) \cup \qquad (2)$$
$$\cup (\overline{P(t-0)} \cdot \overline{P(t)} \cup P(t-0) \cdot P(t)) \cdot (\overline{J(t)} \cdot \overline{K(t)} \cup \overline{J(t)} \cdot \overline{Q(t)} \cup \overline{K(t)} \cdot Q(t))) \cup$$
$$\overline{C(t)} \cdot (\overline{Q(t)} \cdot \overline{P(t-0)} \cdot \overline{P(t)} \cup Q(t) \cdot P(t-0) \cdot P(t)) = 1$$

are called the *equations of the JK flip-flop*. $J, K, C, P, Q$ are signals: the *J input*, the *K input*, the *clock input C*, the *next state P* and the *state Q*.

The first two equations of (1) (modeling the master latch) coincide with the first two equations of the edge triggered RS flip-flop where $S(t) = J(t) \cdot \overline{Q(t)}$, $R(t) = K(t) \cdot Q(t)$ and the last two equations of (1) (modeling the slave latch) coincide with the last two equations of the edge triggered RS flip-flop. We observe that the conditions of admissibility of the inputs of the master and of the slave latch are fulfilled. To be compared (2) and 8 (2).

We have drawn below the JK flip-flop circuit and its symbol:

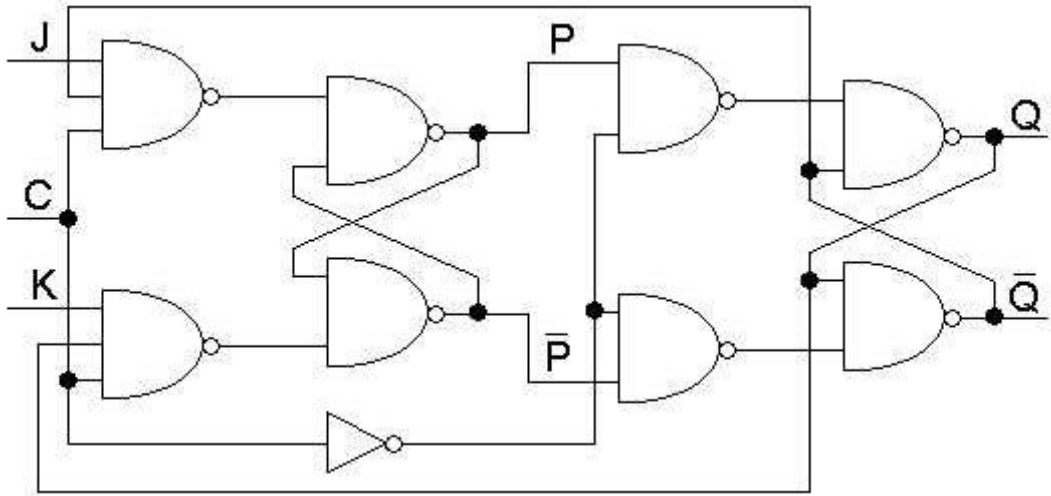

a)

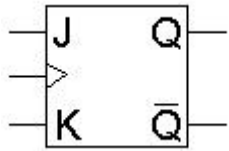

b)

Fig 9

In the analysis of the behavior of this circuit we need to write (2) at the left of $t$ :

$C(t-0) \cdot (P(t-0) \cdot J(t-0) \cdot \overline{Q(t-0)} \cup \overline{P(t-0)} \cdot K(t-0) \cdot Q(t-0) \cup$

$\cup \overline{J(t-0)} \cdot \overline{K(t-0)} \cup \overline{J(t-0)} \cdot \overline{Q(t-0)} \cup \overline{K(t-0)} \cdot Q(t-0)) \cup$

$\overline{C(t-0)} \cdot (\overline{Q(t-0)} \cdot \overline{P(t-0)} \cup Q(t-0) \cdot P(t-0)) = 1$

<u>Case</u> $\overline{C(t-0)} \cdot C(t) = 1$

$\overline{Q(t-0)} \cdot \overline{Q(t)} \cdot \overline{P(t-0)} \cdot (P(t) \cdot J(t) \cup \overline{P(t)} \cdot \overline{J(t)}) \cup$

$Q(t-0) \cdot Q(t) \cdot P(t-0) \cdot (\overline{P(t)} \cdot K(t) \cup P(t) \cdot \overline{K(t)}) = 1$

in other words when $C(t)$ switches from 0 to 1, the next state $P$ and the state $Q$ were equal; $Q$ keeps its value and $P$ switches from 0 to 1 if $J(t) = 1$ and keeps the 0 value otherwise, while the switch of $P$ is from 1 to 0 when $K(t) = 1$, otherwise $P$ remains 1.

<u>Case</u> $C(t) = 1$

$\overline{Q(t-0)} \cdot \overline{Q(t)} \cup Q(t-0) \cdot Q(t) = 1$

$P(t) \cdot J(t) \cdot \overline{Q(t)} \cup \overline{P(t)} \cdot K(t) \cdot Q(t) \cup$

$$\cup \, (\overline{P(t-0)} \cdot \overline{P(t)} \cup P(t-0) \cdot P(t)) \cdot (\overline{J(t)} \cdot \overline{K(t)} \cup \overline{J(t)} \cdot \overline{Q(t)} \cup \overline{K(t)} \cdot Q(t)) = 1$$

$Q$ is constant and the next possibilities exist:

$$J(t) = 0, K(t) = 0$$

$$\overline{P(t-0)} \cdot \overline{P(t)} \cup P(t-0) \cdot P(t) = 1$$

$P$ keeps its value

$$J(t) = 0, K(t) = 1$$

$$\overline{P(t)} \cdot Q(t) \cup (\overline{P(t-0)} \cdot \overline{P(t)} \cup P(t-0) \cdot P(t)) \cdot \overline{Q(t)} = 1$$

The next state is 0 if $Q$ is 1 and keeps its value if $Q$ is 0.

$$J(t) = 1, K(t) = 0$$

$$P(t) \cdot \overline{Q(t)} \cup (\overline{P(t-0)} \cdot \overline{P(t)} \cup P(t-0) \cdot P(t)) \cdot Q(t) = 1$$

The next state is 1 if $Q$ is 0 and keeps its value if $Q$ is 1.

$$J(t) = 1, K(t) = 1$$

$$P(t) \cdot \overline{Q(t)} \cup \overline{P(t)} \cdot Q(t) = 1$$

The next state has a different value from $Q$.

<u>Case</u> $C(t-0) \cdot \overline{C(t)} = 1$

$$Q(t) \cdot P(t-0) \cdot P(t) \cdot (\overline{Q(t-0)} \cup Q(t-0) \cdot \overline{K(t-0)}) \cup$$
$$\overline{Q(t)} \cdot \overline{P(t-0)} \cdot \overline{P(t)} \cdot (Q(t-0) \cup \overline{Q(t-0)} \cdot \overline{J(t-0)}) = 1$$

i.e. the state and the next state become equal in this moment, that is one of continuity for $P$. $Q$ switches from 0 to 1 or keeps the 1 value when $K$ was 0; $Q$ switches from 1 to 0 or keeps the 0 value if $J$ was 0.

<u>Case</u> $\overline{C(t)} = 1$

$$\overline{Q(t)} \cdot \overline{P(t-0)} \cdot \overline{P(t)} \cup Q(t) \cdot P(t-0) \cdot P(t) = 1$$

$P, Q$ are equal and constant.

The JK flip-flop is similar with the edge triggered RS flip-flop, for example $Q$ changes value only when $C(t-0) \cdot \overline{C(t)} = 1$. Let $C(t) = 1$; because $Q(t) = Q(t-0)$ i.e. $Q$ is constant, in the reunion

$$P(t) \cdot J(t) \cdot \overline{Q(t)} \cup \overline{P(t)} \cdot K(t) \cdot Q(t) \cup$$
$$\cup (\overline{P(t-0)} \cdot \overline{P(t)} \cup P(t-0) \cdot P(t)) \cdot (\overline{J(t)} \cdot \overline{K(t)} \cup \overline{J(t)} \cdot \overline{Q(t)} \cup \overline{K(t)} \cdot Q(t))$$

only one of $P(t) \cdot J(t) \cdot \overline{Q(t)}$, $\overline{P(t)} \cdot K(t) \cdot Q(t)$ can be 1, thus $P$ changes value at most once and this was not true at the edge triggered RS flip-flop.

Let's make now in the equations of the $D$ flip-flop $D(t) = J(t) \cdot \overline{Q(t)} \cup \overline{K(t)} \cdot Q(t)$. We get:

$$C(t) \cdot (\overline{Q(t-0)} \cdot \overline{Q(t)} \cup Q(t-0) \cdot Q(t)) \cdot (P(t) \cdot J(t) \cdot \overline{Q(t)} \cup \overline{P(t)} \cdot K(t) \cdot Q(t) \cup \qquad (3)$$
$$\cup \, \overline{P(t)} \cdot \overline{J(t)} \cdot \overline{Q(t)} \cup P(t) \cdot \overline{K(t)} \cdot Q(t)) \cup$$
$$\overline{C(t)} \cdot (\overline{Q(t)} \cdot \overline{P(t-0)} \cdot \overline{P(t)} \cup Q(t) \cdot P(t-0) \cdot P(t)) = 1$$

Equations (2) and (3) have similarities and sometimes the equation of the JK flip-flop is considered to be (3).

## 11. T flip-flop

The next equivalent statements

$$\overline{P(t-0)} \cdot P(t) = \overline{P(t-0)} \cdot \overline{Q(t)} \cdot C(t)$$

$$P(t-0) \cdot \overline{P(t)} = P(t-0) \cdot Q(t) \cdot C(t) \qquad (1)$$

$$\overline{Q(t-0)} \cdot Q(t) = \overline{Q(t-0)} \cdot P(t) \cdot \overline{C(t)}$$

$$Q(t-0) \cdot \overline{Q(t)} = Q(t-0) \cdot \overline{P(t)} \cdot \overline{C(t)}$$

respectively

$$C(t) \cdot (\overline{Q(t-0)} \cdot \overline{Q(t)} \cdot P(t) \cup Q(t-0) \cdot Q(t) \cdot \overline{P(t)}) \cup$$

$$\overline{C(t)} \cdot (\overline{Q(t)} \cdot \overline{P(t-0)} \cdot \overline{P(t)} \cup Q(t) \cdot P(t-0) \cdot P(t)) = 1 \qquad (2)$$

are called the *equations of the T flip-flop*. $C, P, Q$ are signals: the *clock input*, the *next state* and the *state*.

The conditions of admissibility of the inputs are fulfilled for the first two and for the last two equations from (1) (the master and the slave latch).

The T flip-flop circuit and the symbol of the circuit have been drawn below:

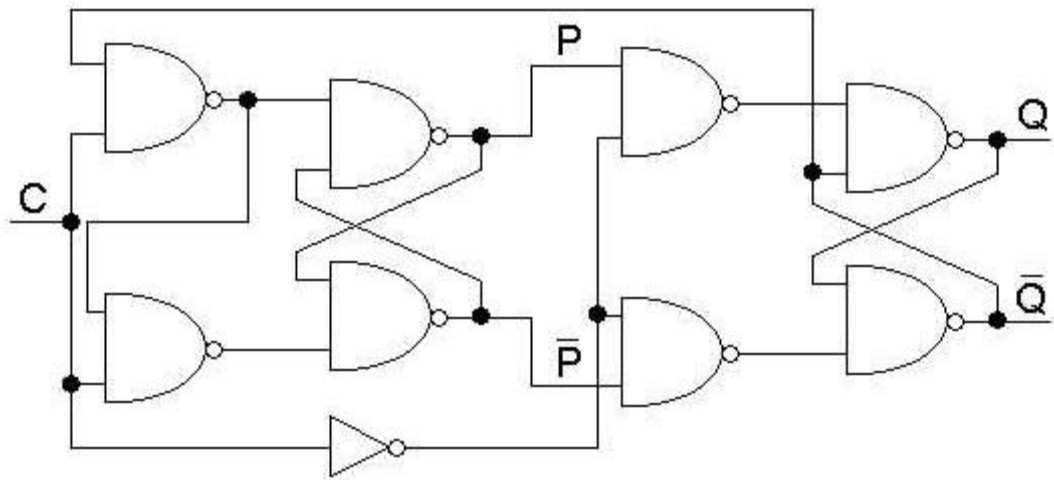

a)

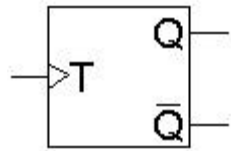

b)

Fig 10

The equation (2) written at the left of $t$ is the next one:

$$C(t-0) \cdot (\overline{Q(t-0)} \cdot P(t-0) \cup Q(t-0) \cdot \overline{P(t-0)}) \cup$$

$$\overline{C(t-0)} \cdot (\overline{Q(t-0)} \cdot \overline{P(t-0)} \cup Q(t-0) \cdot P(t-0)) = 1$$

We interpret the equations of the T flip-flop.

<u>Case</u> $\overline{C(t-0)} \cdot C(t) = 1$

$$\overline{Q(t-0)} \cdot \overline{Q(t)} \cdot \overline{P(t-0)} \cdot P(t) \cup Q(t-0) \cdot Q(t) \cdot P(t-0) \cdot \overline{P(t)} = 1$$

the state and the next state were equal, but now they have complementary values; $t$ is a point of continuity for $Q$ and of discontinuity for $P$.

<u>Case</u> $C(t) = 1$

$$\overline{Q(t-0)} \cdot \overline{Q(t)} \cdot P(t) \cup Q(t-0) \cdot Q(t) \cdot \overline{P(t)} = 1$$

$P, Q$ are constant, with complementary values.

<u>Case</u> $C(t-0) \cdot \overline{C(t)} = 1$

$$\overline{Q(t-0)} \cdot Q(t) \cdot P(t-0) \cdot P(t) \cup Q(t-0) \cdot \overline{Q(t)} \cdot \overline{P(t-0)} \cdot \overline{P(t)} = 1$$

the state and the next state were different and they become equal; $t$ is a point of continuity for $P$ and of discontinuity for $Q$.

<u>Case</u> $\overline{C(t)} = 1$

$$\overline{Q(t)} \cdot \overline{P(t-0)} \cdot \overline{P(t)} \cup Q(t) \cdot P(t-0) \cdot P(t) = 1$$

$P, Q$ are constant and equal.

At each falling edge $C(t-0) \cdot \overline{C(t)} = 1$ of the clock input, the state $Q$ of the T flip-flop toggles to its complementary value, otherwise it is constant.

We observe that the equations of the T flip-flop represent the next special cases:

- in the equations of the edge triggered RS flip-flop, $S(t) = \overline{Q(t)}, R(t) = Q(t)$
- in the equations of the D flip-flop $D(t) = \overline{Q(t)}$
- in the equations of the JK flip-flop (any of 10 (2), 10 (3)) $J(t) = 1, K(t) = 1$

## Appendix

We prove that the statements

$$\overline{x(t-0)} \cdot x(t) = \overline{x(t-0)} \cdot u(t)$$
$$x(t-0) \cdot \overline{x(t)} = x(t-0) \cdot v(t) \qquad (1)$$
$$u(t) \cdot v(t) = 0$$

and

$$x(t) \cdot u(t) \cdot \overline{v(t)} \cup \overline{x(t)} \cdot \overline{u(t)} \cdot v(t) \cup (\overline{x(t-0)} \cdot \overline{x(t)} \cup x(t-0) \cdot x(t)) \cdot \overline{u(t)} \cdot \overline{v(t)} = 1 \quad (2)$$

are equivalent. (1) is equivalent with (we have underlined some terms that are reduced):

$$1 = (\overline{\overline{x(t-0)} \cdot x(t)} \cdot \overline{\overline{x(t-0)} \cdot u(t)} \cup \overline{x(t-0)} \cdot x(t) \cdot \overline{\overline{x(t-0)} \cdot u(t)}) \cdot$$
$$\cdot (\overline{x(t-0) \cdot \overline{x(t)}} \cdot \overline{x(t-0) \cdot v(t)} \cup x(t-0) \cdot \overline{x(t)} \cdot x(t-0) \cdot v(t)) \cdot$$
$$\cdot (\overline{u(t)} \cup \overline{v(t)}) =$$
$$= ((x(t-0) \cup \overline{x(t)}) \cdot (x(t-0) \cup \overline{u(t)}) \cup \overline{x(t-0)} \cdot x(t) \cdot u(t)) \cdot$$
$$\cdot ((\overline{x(t-0)} \cup x(t)) \cdot (\overline{x(t-0)} \cup \overline{v(t)}) \cup x(t-0) \cdot \overline{x(t)} \cdot v(t)) \cdot$$
$$\cdot (\overline{u(t)} \cup \overline{v(t)}) =$$
$$= (x(t-0) \cup \underline{x(t)} \cdot \overline{u(t)} \cup x(t-0) \cdot \overline{x(t)} \cup \overline{x(t)} \cdot \overline{u(t)} \cup \overline{x(t-0)} \cdot x(t) \cdot u(t)) \cdot$$
$$\cdot (\overline{x(t-0)} \cup \overline{x(t-0)} \cdot \overline{v(t)} \cup \underline{x(t-0)} \cdot x(t) \cup x(t) \cdot \overline{v(t)} \cup x(t-0) \cdot \overline{x(t)} \cdot v(t)) \cdot$$
$$\cdot (\overline{u(t)} \cup \overline{v(t)}) =$$
$$= (x(t-0) \cup \overline{x(t)} \cdot \overline{u(t)} \cup \overline{x(t-0)} \cdot x(t) \cdot u(t)) \cdot$$
$$\cdot (\overline{x(t-0)} \cup x(t) \cdot \overline{v(t)} \cup x(t-0) \cdot \overline{x(t)} \cdot v(t)) \cdot$$
$$\cdot (\overline{u(t)} \cup \overline{v(t)}) =$$
$$= (\overline{x(t-0)} \cdot \overline{x(t)} \cdot \overline{u(t)} \cup \overline{x(t-0)} \cdot x(t) \cdot u(t) \cup$$
$$\cup x(t-0) \cdot x(t) \cdot \overline{v(t)} \cup \overline{x(t-0)} \cdot x(t) \cdot u(t) \cdot \overline{v(t)} \cup$$
$$\cup x(t-0) \cdot \overline{x(t)} \cdot v(t) \cup x(t-0) \cdot \overline{x(t)} \cdot \overline{u(t)} \cdot v(t)) \cdot$$

$$\cdot \, (\overline{u(t)} \cup \overline{v(t)}) =$$

$$= \overline{(x(t-0) \cdot \overline{x(t)} \cdot \overline{u(t)}} \cup \overline{x(t-0)} \cdot x(t) \cdot u(t) \cup x(t-0) \cdot x(t) \cdot \overline{v(t)} \cup$$

$$\cup \, x(t-0) \cdot \overline{x(t)} \cdot v(t)) \cdot (\overline{u(t)} \cup \overline{v(t)}) =$$

$$= \overline{x(t-0) \cdot \overline{x(t)} \cdot \overline{u(t)}} \cup x(t-0) \cdot x(t) \cdot \overline{u(t)} \cdot \overline{v(t)} \cup x(t-0) \cdot \overline{x(t)} \cdot \overline{u(t)} \cdot v(t) \cup$$

$$\cup \, \overline{x(t-0) \cdot \overline{x(t)} \cdot \overline{u(t)} \cdot v(t)} \cup \overline{x(t-0)} \cdot x(t) \cdot u(t) \cdot \overline{v(t)} \cup x(t-0) \cdot x(t) \cdot \overline{v(t)} =$$

$$= \overline{x(t-0) \cdot \overline{x(t)} \cdot \overline{u(t)}} \cdot (v(t) \cup \overline{v(t)}) \cup$$

$$\cup \, x(t-0) \cdot x(t) \cdot \overline{u(t)} \cdot \overline{v(t)} \cup x(t-0) \cdot \overline{x(t)} \cdot \overline{u(t)} \cdot v(t) \cup$$

$$\cup \, \overline{x(t-0) \cdot \overline{x(t)} \cdot \overline{u(t)} \cdot v(t)} \cup \overline{x(t-0)} \cdot x(t) \cdot u(t) \cdot \overline{v(t)} \cup$$

$$\cup \, x(t-0) \cdot x(t) \cdot (u(t) \cup \overline{u(t)}) \cdot \overline{v(t)} =$$

$$= \overline{x(t-0) \cdot \overline{x(t)} \cdot \overline{u(t)}} \cdot v(t) \cup \overline{x(t-0) \cdot \overline{x(t)} \cdot \overline{u(t)}} \cdot \overline{v(t)} \cup$$

$$\cup \, x(t-0) \cdot x(t) \cdot \overline{u(t)} \cdot \overline{v(t)} \cup x(t-0) \cdot \overline{x(t)} \cdot \overline{u(t)} \cdot v(t) \cup$$

$$\cup \, \overline{x(t-0) \cdot \overline{x(t)} \cdot \overline{u(t)} \cdot v(t)} \cup \overline{x(t-0)} \cdot x(t) \cdot u(t) \cdot \overline{v(t)} \cup$$

$$\cup \, x(t-0) \cdot x(t) \cdot u(t) \cdot \overline{v(t)} \cup \overline{x(t-0)} \cdot x(t) \cdot \overline{u(t)} \cdot \overline{v(t)} =$$

$$= x(t) \cdot u(t) \cdot \overline{v(t)} \cdot (\overline{x(t-0)} \cup x(t-0)) \cup$$

$$\cup \, \overline{x(t)} \cdot \overline{u(t)} \cdot v(t) \cdot (\overline{x(t-0)} \cup x(t-0)) \cup$$

$$\cup \, (\overline{x(t-0)} \cdot \overline{x(t)} \cup x(t-0) \cdot x(t)) \cdot \overline{u(t)} \cdot \overline{v(t)} =$$

$$= x(t) \cdot u(t) \cdot \overline{v(t)} \cup \overline{x(t)} \cdot \overline{u(t)} \cdot v(t) \cup (\overline{x(t-0)} \cdot \overline{x(t)} \cup x(t-0) \cdot x(t)) \cdot \overline{u(t)} \cdot \overline{v(t)}$$